\documentclass[conference]{IEEEtran}

\usepackage{tikz}
\usepackage{standalone}
\usepackage[T1]{fontenc}
\usepackage{amsmath}
\usepackage[cmintegrals]{newtxmath}
\usepackage{pgfplots}
\pgfplotsset{compat=1.14}
\usepackage{algorithm}
\newlength{\hsep}

\newcommand{\define}{\stackrel{\Delta}{=}}

\def\So{\mathcal{S}_{\circ}}
\def\calA{\mathcal{A}}

\def\calS{\mathcal{S}}

\def\calE{\mathcal{E}}
\def\calX{\mathcal{X}}

\def\rloss{R_{\text{loss}}}

\def\emax{E_{\max}}

\def\Gs{\text{G}_{\text{s}}}

\def\rloss{R_{\text{loss}}}

\def\snr{\text{SNR}}

\def\exp{\mathbb{E}}
\def\ent{\mathbb{H}}

\def\rbmd{\text{R}_{\text{BMD}}}
\def\capc{C}
\def\delsnr{\Delta\text{SNR}}

\colorlet{shapercolor}{green!20!white}
\colorlet{bshapercolor}{cyan!20!white}
\colorlet{feccolor}{blue!20!white}
\colorlet{mapcolor}{red!20!white}
\colorlet{chancolor}{black!10!white}

\def\lc{\left\lceil}   
\def\rc{\right\rceil}
\def\lf{\left\lfloor}   
\def\rf{\right\rfloor}

\usetikzlibrary{plotmarks,matrix,chains,scopes,fit,calc,shapes,positioning,decorations,intersections,arrows,backgrounds,shadows}

\newlength\FigureHeight
\newlength\FigureWidth
\begin{document}

\title{Partial Enumerative Sphere Shaping}
\author{\IEEEauthorblockN{Yunus Can G\"{u}ltekin\IEEEauthorrefmark{1},
W. J. van Houtum\IEEEauthorrefmark{2},
Arie Koppelaar\IEEEauthorrefmark{3} and 
Frans M. J. Willems\IEEEauthorrefmark{1} }
\IEEEauthorblockA{\IEEEauthorrefmark{1}Eindhoven University of Technology, Eindhoven, The Netherlands}
\IEEEauthorblockA{\IEEEauthorrefmark{2}Catena Radio Design, Eindhoven, The Netherlands}
\IEEEauthorblockA{\IEEEauthorrefmark{3}NXP Semiconductors, Eindhoven, The Netherlands} Email: y.c.g.gultekin@tue.nl }

\maketitle

\begin{abstract}
The dependency between the Gaussianity of the input distribution for the additive white Gaussian noise (AWGN) channel and the gap-to-capacity is discussed.
We show that a set of particular approximations to the Maxwell-Boltzmann (MB) distribution virtually closes most of the shaping gap.
We relate these symbol-level distributions to bit-level distributions, and demonstrate that they correspond to keeping some of the amplitude bit-levels uniform and independent of the others.
Then we propose partial enumerative sphere shaping (P-ESS) to realize such distributions in the probabilistic amplitude shaping (PAS) framework.
Simulations over the AWGN channel exhibit that shaping 2 amplitude bits of 16-ASK have almost the same performance as shaping 3 bits, which is 1.3 dB more power-efficient than uniform signaling at a rate of 3 bit/symbol.
In this way, required storage and computational complexity of shaping are reduced by factors of 6 and 3, respectively.
\end{abstract}

\IEEEpeerreviewmaketitle

\section{Introduction}
\begin{figure}[ht]
\includegraphics[width=\columnwidth]{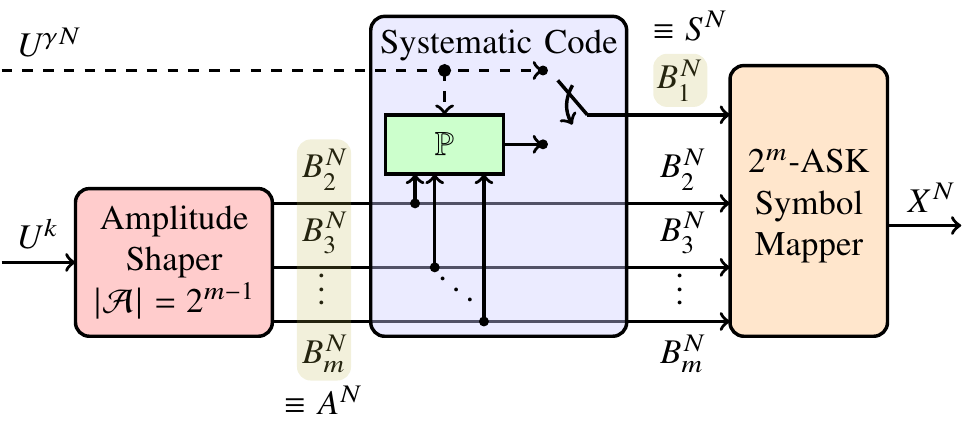}
\caption{Probabilistic amplitude shaping transmitter.}
\label{fig:PASblockdiag}
\end{figure}

The probabilistic amplitude shaping (PAS) architecture is introduced in~\cite{BochererSS2015_ProbAmpShap} to integrate amplitude shaping with forward error correction (FEC) to close the shaping gap.
The idea is to realize shaping over the amplitudes of the channel inputs.
Then error correction is achieved by coding the signs based on the binary labels of these amplitudes, see Fig.~\ref{fig:PASblockdiag}.
The corner stone of the PAS construction is the amplitude shaping block which maps uniform bits to shaped amplitude sequences that satisfy a predefined condition.
To realize this transformation in an invertible manner, different types of amplitude shaping algorithms have been proposed: Constant composition distribution matching (CCDM)~\cite{SchulteB2016_CCDM}, multiset-partition distribution matching (MPDM)~\cite{Fehenberger2019_MPDM}, enumerative sphere shaping (ESS)~\cite{WillemsW1993_ESS}, shell mapping (SM)~\cite{Schulte2019_SMDM}, etc.

Motivated by the observation that $N$-spherical signal structures lead to Gaussian distributed inputs asymptotically for large $N$~\cite{Forney1984_EffModBLchan}, ESS, i.e., sphere shaping, utilizes amplitude sequences satisfying a maximum-energy constraint.
Since all signal points inside a sphere are employed, ESS minimizes the rate loss for a given shaping rate at any dimension~\cite{Gultekin2018_ConstShapforShortBlocks}.
Furthermore, ESS requires significantly smaller computational complexity than SM which is another method for realizing sphere shaping.
Thus, it is a shaping technique which is suitable for short block length applications and is considered in the PAS framework for wireless communications in~\cite{GultekinHKW2019_ESSforShortWlessComm}. 

In the context of binary FEC and bit-metric decoding (BMD), shaping the amplitudes of the channel inputs creates shaped bit-levels which are dependent on each other in general.
As another approach, product distribution matching (PDM) is proposed to keep bit-levels independent of each other while shaping them using multiple binary distribution matchers~\cite{SteinerSB2018_PDM}.
A brief review of symbol-level and bit-level probabilistic shaping strategies can be found in~\cite[Sec. II]{BochererSS2015_ProbAmpShap}.

In this work, we first show through a gap-to-capacity analysis that most of the maximum capacity gain can be obtained without shaping all amplitude bit-levels.
Then we propose Partial ESS (P-ESS) as a solution to shape only a subset of the amplitude bit-levels. 
The basic principle is to use ESS with an amplitude alphabet with smaller cardinality than that of the original system which decreases required storage and computational complexity of shaping.
By combining the shaped bit-levels produced by the shaper with data bits as the uniform levels, we obtain a {\it partially-shaped} constellation that provides a performance quite close to a fully-shaped constellation.

Furthermore, there is a lower bound on the rate of the FEC code to be used in the PAS architecture by design~\cite[Sec. IV-D]{BochererSS2015_ProbAmpShap}.
This bound increases with increasing constellation size.
If P-ESS is used instead of symbol-level ESS, we can relax this constraint which is important especially for very large constellation sizes.

The current paper is structured as follows.
In Sec.~\ref{background}, some background information on amplitude shaping is provided.
In Sec.~\ref{sec:gap2capacity}, we investigate approximate Maxwell-Boltzmann distributions from a gap-to-capacity perspective.
How to realize these distributions based on enumerative sphere shaping is explained in Sec.~\ref{sec:pess}.
Subsequently, we provide numerical results in Sec.~\ref{sec:results} before concluding the paper.

\section{Background on Amplitude Shaping}
\label{background}
\subsection{AWGN Capacity}
The discrete-time AWGN channel output is modeled at time $n = 1, 2,\cdots, N$ as $Y_n = X_n + Z_n$ where $X_n$ and $Y_n$ are the channel input and output, respectively\footnote{Notation: Random variables and random vectors are denoted by $X$ and $X^N$, respectively.
Realizations of them are indicated by $x$ and $x^N$, respectively. $P_X(x)$ denotes the probability mass function (PMF) for $X$.}.
Noise $Z_n$ is drawn from a zero-mean Gaussian distribution with variance $\sigma^2$ and independent of $X_n$.
The block length is specified by $N$ in real symbols.
There is an average input power constraint, i.e., $\exp[X^2]\leq P$, where $\exp$ denotes the expectation operator.
The signal-to-noise ratio (SNR) is defined as $\snr=\exp[X^2]/\sigma^2$.
The capacity of this channel, $C=\frac{1}{2}\log_2(1+\snr)$, is achieved only when $X$ is a zero-mean Gaussian with variance $P$~\cite{CoverT1991_ElementsofInfoTheo}.
The corresponding random coding argument shows that input sequences, drawn from a Gaussian distribution, are likely to lie in an $N$-sphere of squared radius $N(P+\varepsilon)$ when $N\rightarrow \infty$, for small positive $\varepsilon$. Thus, it makes sense to use an $N$-sphere as the signal space boundary, to achieve the capacity $C$.
\vspace{-0.08cm}

\subsection{Amplitude Shaping for Discrete Constellations}
We consider $M$-ASK that results in the alphabet $\calX = \{\pm 1, \pm 3,\cdots, \pm (2^{m}-1)\}$ for $m\geq2$ where $M=2^m$.
This alphabet can be factorized as $\calX = \calS\times\calA$ where $\calS = \{-1, 1\}$ and $\calA = \{1, 3,\cdots, 2^{m}-1\}$ are the sign and amplitude alphabets, respectively.
There is no analytical expression for the distribution that maximizes the achievable information rate (AIR) for an ASK constellation.
For such constellations, Maxwell-Boltzmann (MB) distributions~\cite[Sec. III-C]{BochererSS2015_ProbAmpShap}
\begin{equation}
P_{\text{MB}}(a) = K\left(\lambda\right) e^{-\lambda a^2},\mbox{ $a \in \calA$}, \label{eq:mbdist}
\end{equation}
are pragmatically chosen for amplitude shaping~\cite{BochererSS2015_ProbAmpShap,Kschischang1993_OptimalNonUnifSignaling}, since they maximize the rate for a fixed average power, or equivalently, minimize the average energy for a given entropy~\cite{CoverT1991_ElementsofInfoTheo}.
In a dual manner, sphere shaping of multidimensional ASK constellations is realized to create Gaussian-like-distributed channel inputs~\cite{WillemsW1993_ESS,LaroiaFT1994_OptimalShaping}.
We emphasize that although MB-distributed and sphere-shaped ASK constellations maximize the energy efficiency, they do not maximize the AIR.
\vspace{-0.08cm}

\subsection{Binary Labeling} \label{ssec:binlab}
To combine higher order modulation with binary FEC, a binary labeling strategy is needed.
We assume that the binary label $B_1B_2\cdots B_{m}$ of a $2^{m}$-ASK constellation point $X$ can be decomposed into the amplitude bit-levels $B_2B_3\cdots B_{m}$ and the sign bit-level $B_{1}$.
The amplitude part of a binary reflected Gray code (BRGC) for 16-ASK is provided in Table~\ref{tab:16askBRGC}.

\begin{table}[ht]
\begin{center}
\caption{}
\resizebox{0.75\columnwidth}{!}{
\begin{tabular}{c|cccccccc} 
$A$ & 1 & 3 & 5 & 7 & 9 & 11 & 13 & 15 \\
\hline\hline
$B_2$ & 0 & 0 & 0 & 0 & 1 & 1 & 1 & 1 \\ 
$B_3$ & 0 & 0 & 1 & 1 & 1 & 1 & 0 & 0 \\ 
$B_4$ & 0 & 1 & 1 & 0 & 0 & 1 & 1 & 0 \\ 
\end{tabular}
}
\label{tab:16askBRGC}
\end{center}
\end{table}

\subsection{Probabilistic Amplitude Shaping} \label{sec:pas}
The probabilistic amplitude shaping (PAS) architecture is proposed in~\cite{BochererSS2015_ProbAmpShap} to provide integration of shaping into existing FEC schemes.
At the transmitter, see Fig.~\ref{fig:PASblockdiag}, a $k$-bit uniform data sequence $U^k$ is mapped to a shaped amplitude sequence $A^N$.
Then the binary label sequences $B_2^NB_3^N\cdots B_m^N$ of these amplitudes are passed to a systematic FEC code of rate $R_c=(m-1)/m$.
The $N$ parity bits produced by the encoder are used to specify the signs $S^N$.
As shown in Fig.~\ref{fig:PASblockdiag}, this can be accomplished using a parity-check matrix $\mathbb{P}$.
Finally, $X^N=S^N\times A^N$ is transmitted over the channel.
The transmission rate of this scheme is $R_t = k/N$~bits per real dimension (bit/1-D).

If a FEC code of rate $R_c>(m-1)/m$ is employed, an extra $\gamma N$-bit uniform information sequence $U^{\gamma N}$ is used as the input of the encoder along with $B_2^NB_3^N\cdots B_m^N$. 
This modified architecture has the dashed branches activated in Fig.~\ref{fig:PASblockdiag}.
Now,  both the additional data bits $U^{\gamma N}$ and the parity output of the encoder are used as the sign bit-level $B_{1}$.
The fraction of signs that are specified by the extra data is denoted by $\gamma = R_cm-(m-1)$.
The transmission rate of this scheme is $R_t = k/N + \gamma$~bit/1-D.

Imposing a non-uniform distribution over the amplitudes creates shaped amplitude bit-levels $B_2B_3\cdots B_m$ which are dependent on each other in general.
We call such schemes fully-shaped or $(m-1)$-bit shaped since all amplitude bit-levels are stochastically affected.

\subsection{Finite Length Rate Loss} \label{ssec:rloss}
We define the rate loss of a finite length shaping technique of shaping rate $R_s$~bits per amplitude (bit/amp.) as~\cite{GultekinHKW2019_ESSforShortWlessComm}:
\begin{eqnarray}
\rloss \define \ent(A_{\text{MB}})-R_s, \label{eq:rloss}
\end{eqnarray}
where $A_{\text{MB}}$ is the MB-distributed random variable with expected symbol energy $\sum_{a\in\calA} a^2 P_A(a)$, and $P_A(a)$ is the induced distribution over the amplitudes $a\in\calA$ due to shaping.
In~\cite{Gultekin2018_ConstShapforShortBlocks}, we showed that there exist shaping codes with vanishing $\rloss$ as $N\rightarrow\infty$.
We use the rate loss as a performance indicator for shaping codes at finite block lengths.

\subsection{Achievable Information Rate} \label{ssec:air}
At the receiver, log-likelihood ratios (LLR) of each bit-level are computed by a soft demapper.
Note that the non-uniform a priori distribution of the amplitudes is taken into account, see~\cite[Sec. VI]{BochererSS2015_ProbAmpShap}.
Then a bit-metric decoder uses these LLRs to retrieve the transmitted information.
The BMD rate~\cite{Bocherer2014_ProbSigShapForBMD} 
\begin{eqnarray}
\rbmd &=& \left[ \mathbb{H}(X) - \sum_{i=1}^{m} \mathbb{H}(\text{B}_i|Y ) \right]^+, \label{rbmd}
\end{eqnarray}
is achievable by a bit-metric decoder for any input distribution $P_{X}(x)$ where $[\cdot]^+ = \max\{0,\cdot\}$.

\section{`Gaussianity' and Gap-to-Capacity} \label{sec:gap2capacity}
\subsection{Maxwell-Boltzmann-Distributed Inputs} \label{ssec:gap2capacity}
\begin{figure}[ht]
\centering
\includegraphics[width=0.91\columnwidth]{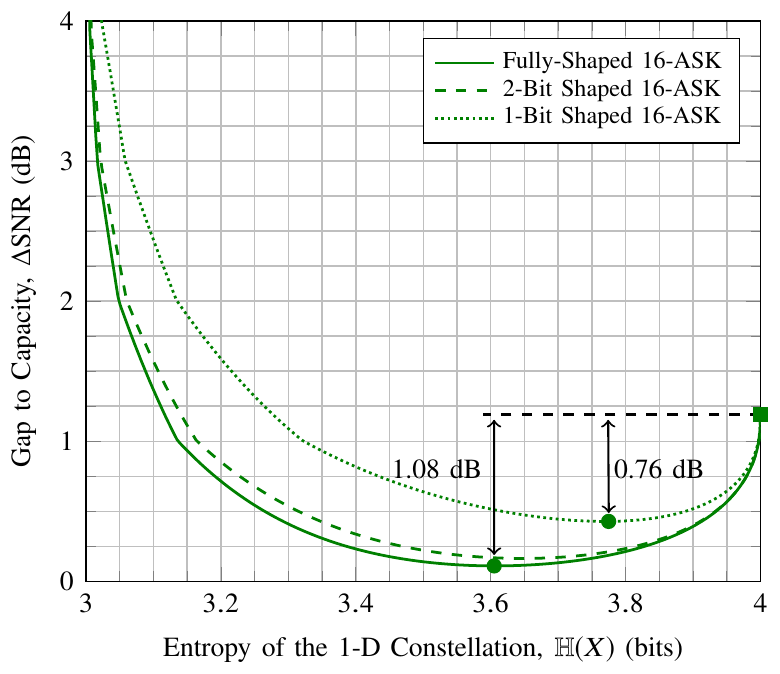}
\caption{Constellation entropy vs. gap-to-capacity for 16-ASK at rate $R_t = 3$.}
\label{fig:g2c}
\end{figure}

To obtain a transmission rate $R_t$ employing the $2^m$-ASK constellation, a total $m-R_t$~bits redundancy is added in the PAS construction.
Shaping and FEC coding are responsible for $m-\ent(X)$ and $\ent(X)-R_t$~bits of the total redundancy, respectively.
Thus, provided that $A$ is MB-distributed, the input entropy $\ent(X)=\ent(A)+1$ is a design choice and can be used to adjust the balance between the shaping and coding redundancies.
Following Wachsmann et al.~\cite[Sec. VIII]{Wachsmann1999_MLcodes}, we minimize gap-to-capacity
\begin{equation}
\delsnr = 10\log \left( \snr \right) \bigg|_{\rbmd=R_t} - 10\log \left( \snr \right) \bigg|_{\capc=R_t}, \label{eq:g2c}
\end{equation}
to find the optimum MB-distributed input entropy $\ent(X)$.
In Fig.~\ref{fig:g2c}, $\delsnr$ is plotted versus $\ent(X)$ at the target rate $R_t = 3$ with the solid line.
Here, $A$ is assumed to be MB-distributed over the 16-ASK amplitude alphabet.
For a set of $\lambda$ values, see \eqref{eq:mbdist}, $\rbmd$ is computed, using the labeling given in Table~\ref{tab:16askBRGC}.

The rightmost point of the curves in Fig.~\ref{fig:g2c} belongs to uniform signaling where the target rate is obtained using a code of rate $R_c = R_t/m = 3/4$.
The leftmost part corresponds to uncoded signaling and the target rate is achieved by shaping the constellation such that $\ent(X) = R_t = 3$.
Here $\delsnr$ is infinite since without channel coding, error-free communication is only possible over a noiseless channel.

In Fig.~\ref{fig:g2c}, the gap-to-capacity is minimized for 3-bit shaped 16-ASK at $\ent(X)=3.63$.
This suggests that the shaping operation should add 0.37 bits of redundancy while coding adds 0.63 bits.
At this optimum point, the gain over uniform signaling is 1.08 dB.
The corresponding FEC code rate for the PAS structure is $R_c = 0.84$, see~\cite[(30)]{GultekinHKW2019_ESSforShortWlessComm}.

\subsection{Partially Maxwell-Boltzmann-Distributed Inputs} \label{ssec:approxmb}
To investigate how well the amplitude distribution $P_A(a)$ has to resemble a MB distribution to close most of the shaping gap, we define a particular type of approximation.
We will later relate these approximate symbol-level distributions to bit-level distributions.
The basic idea here is to realize MB distributions over amplitude pairs, quartets, etc., instead of individual amplitudes.
We now explain this with an example.

\subsubsection*{\bf Example (Partially MB-distributed 16-ASK)}
We consecutively gather amplitudes of 16-ASK alphabet into groups of two, i.e., $\calA_{1}=\{1,3\}$, $\calA_{2}=\{5,7\}$, $\calA_{3}=\{9,11\}$ and $\calA_{4}=\{13,15\}$.
Then we define the MB distribution, see \eqref{eq:mbdist}, over these pairs as
\begin{equation}
P_{\text{MB}}\left(a\in\calA_i\right) = K\left(\lambda\right) e^{-\lambda \exp\left[\left| \calA_{i} \right|^2\right]},\label{eq:approxmb}
\end{equation}
where $\exp[|\calA_{i}|^2]$ is the average energy of $a \in \calA_i$ assuming they are equiprobable, more precisely, $\exp [|\calA_{i}|^2] = \frac{1}{2}\sum_{a \in \calA_i} |a|^2$ for $i \in \{1,2,3,4\}$.
While $\lambda$ can be adjusted to fix $\mathbb{H}(A)$, this approximation can be realized over symbol quartets $\{1, 3, 5, 7\}$ and $\{9, 11, 13, 15\}$, and so on.

\begin{table}[ht]
\caption{}
\renewcommand{\arraystretch}{1.3}
\centering
\resizebox{\columnwidth}{!}{
\begin{tabular}{cccccccccc} 
\hline
$p_1$ & $p_3$ & $p_5$ & $p_7$ & $p_9$ & $p_{11}$ & $p_{13}$ & $p_{15}$ & $E$ & $\Gs$ \\
\hline
.2443 & .2225 & .1847 & .1396 & .0962 & .0603 & .0345 & .0180 & 38.66 & 1.40\\
.2365 & .2365 & .1623 & .1623 & .0765 & .0765 & .0247 & .0247 & 39.57 & 1.30\\ 
.2065 & .2065 & .2065 & .2065 & .0435 & .0435 & .0435 & .0435 & 43.27 & 0.92\\
\hline  
\end{tabular}
}
\label{tab:probabilities}
\end{table}

In Table~\ref{tab:probabilities}, a numeric example is tabulated for $P_A(a)=p_a$ for 16-ASK where $\mathbb{H}(A)=2.667$.
Here the first row is the exact MB distribution and the following are the approximations over 2- and 4-symbol groups, respectively, all rounded to the nearest 4 decimal digits.
Average symbol energy $E=\exp[A^2]$ and the shaping gain $\Gs$ (in dB) with respect to uniform signaling as in~\cite[(10)]{GultekinHKW2019_ESSforShortWlessComm} are also provided in Table~\ref{tab:probabilities}.
We see that as we apply the MB distribution over symbols, pairs and quartets for a fixed entropy, $E$ increases which indicates that energy efficiency is decreasing.
This can also be verified by observing the decreasing shaping gain.

When MB distributions are realized over symbol pairs or quartets, 1 or 2 amplitude bit-levels become uniform and independent of the others, respectively.
Considering the example above, pairing the amplitudes of 16-ASK and transmitting the elements of a group equiprobably means that the bit-level $B_4$ is now uniform and independent of the other two amplitude bits, see Table.~\ref{tab:16askBRGC}.
Similarly, assigning the same probability to each amplitude in a group of four implies that bit-levels $B_3B_4$ are uniform and independent of the others.
We call these schemes $s$-bit shaped where $s<m-1$ is the number of shaped amplitude bit-levels.

In Fig.~\ref{fig:g2c}, gap-to-capacity is also plotted for 2- and 1-bit shaped 16-ASK.
The important observation is that it is possible to obtain a large portion of the maximum possible gain even when only some of the amplitude bits are shaped.
As an example, the maximum gain for 16-ASK at $R_t = 3$ drops from 1.08 dB to 1.03, and then to 0.76 when 2 and 1 bits are shaped instead of 3, respectively, see Fig.~\ref{fig:g2c}.
We note here that a similar gap-to-capacity analysis is provided for product distributions\footnote{`Product distribution' specifies symbol-level distributions in the form of the product of bit-level distributions~\cite{SteinerSB2018_PDM}.} in~\cite{SteinerSB2018_PDM}.
Next, we will build an amplitude shaping block to realize output distributions resembling the approximate distributions in \eqref{eq:approxmb}.

\section{Partial Enumerative Sphere Shaping} \label{sec:pess}
\subsection{Enumerative Sphere Shaping} \label{ssec:ess}
Let $\So$ denote the set of amplitude sequences $a^N$ having energy not larger than $\emax$, i.e., $\sum_{j=1}^{N} a_j^2 \leq \emax$ where $a_j \in \calA$ for $j = 1, 2, \cdots, N$. 
This set specifies an $N$-spherical region of the amplitude lattice $\calA^N$ in $N$-dimensional space\footnote{Here we use $\calA^N$ to denote the $N$-fold Cartesian product of $\calA$ with itself.}.
Enumerative sphere shaping (ESS) creates an invertible mapping from integer message indices $i \in [0,|\So|)$ to amplitude sequences $a^N \in \So$ by ordering them lexicographically~\cite{GultekinHKW2019_ESSforShortWlessComm}. 

\begin{figure}[ht]
\centering
\includegraphics[width=\columnwidth]{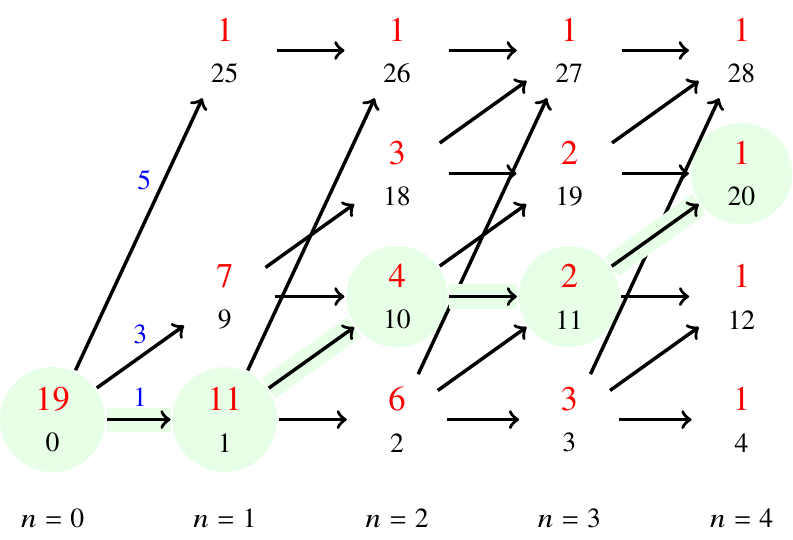}
\caption{ESS trellis constructed for $N=4$, $\calA = \{1, 3, 5, 7\}$ and $\emax=28$.}
\label{esstrellis}
\end{figure}

For this indexing purpose, a trellis is constructed.
As an example, consider the trellis given in Fig.~\ref{esstrellis} which is created with $N=4$, $\calA = \{1, 3, 5, 7\}$ and $\emax=28$. 
Here, each $a^N\in\So$ is represented by a path composed of $N$ branches, e.g., the sequence $(1,3,1,3)$ is highlighted by light green in Fig.~\ref{esstrellis}.
Each branch depicts an amplitude from $\calA$ which are notated by small blue letters.
The node that a path travels through in $n^{\text{th}}$ column identifies its accumulated energy over the first $n$ dimensions, i.e., $\sum_{j=1}^{n} a_j^2$.
Any node can be identified by the dimension-energy pair $(n,e)$.
The energy values are indicated by small black letters.
Each path starts from the zero-energy node and ends in a node from the $N^{\text{th}}$ column. 
The nodes in the last column represent possible sequence energies that take values from $\left\{ N, N+8, N+16,\cdots, \emax \right\}$ for ASK constellations.
The number of energy levels is $L = \lf (\emax-N)/8 \rf +1$.
Equivalently, $L$ is the number of $N$-dimensional shells that the signal points are located on.

The larger red numbers in Fig.~\ref{esstrellis} specify the number of paths $T_n^e$ that lead from node $(n,e)$ to a final node for $n = 0, 1,\cdots, N-1$ and $e\leq\emax$. 
Thus $T_0^0=|\So|$ is the number of sequences represented in the trellis, and $R_s = \log_2|\So|/N$ is the shaping rate of the corresponding sphere code in bit/amp.
Values of $T_n^e$ can be computed recursively for $n=N-1, N-2,\cdots, 0$, after initializing the last column with ones, i.e., $T_N^e=1$, and stored as a matrix~\cite{GultekinHKW2019_ESSforShortWlessComm}.
Note that in column $n$, we only consider states with possible energy levels between $n$ and  $\emax +n -N$.
Enumerative shaping and deshaping algorithms use $T_n^e$ to compute the sequence with a given index, and vice versa~\cite{GultekinHKW2019_ESSforShortWlessComm}.

\subsection{Implementation Aspects of ESS} \label{ssec:implementation}
In the PAS architecture, the function of ESS is to map uniform binary data sequences to shaped amplitude sequences.
The input length of an enumerative sphere shaper is $k = \lf \log_2 |\So| \rf$~bits.
Thus, only the sequences with indices smaller than $2^k$ are actually transmitted.

To store the trellis, at most $L (N+1) \lc NR_s \rc$~bits of memory is required.
The computational complexity of the indexing algorithms is $(|\calA|-1)\lc NR_s \rc$ bit operations per dimension (bit oper./1-D)~\cite{GultekinHKW2019_ESSforShortWlessComm}.
The trellis can also be computed with bounded-precision where each number $T_n^e$ is rounded down, i.e., approximated, as $T_n^e \approx m\cdot 2^p$.
Here $m$ and $p$ are called the mantissa and the exponent which are stored using $n_m$ and $n_p$ bits, respectively.
The storage and computational complexities of this bounded-precision trellis computation and the corresponding indexing algorithms are $L (N+1) (n_m+n_p)$~bits and $(|\calA|-1)(n_m+n_p)$~bit oper./1-D, respectively~\cite{GultekinWHS2018_ApproxEnumerative}.

\subsection{Partial Enumerative Sphere Shaping} \label{ssec:pess}
We want to transmit channel inputs drawn from the $2^m$-ASK alphabet, and we want the distribution of their amplitudes resemble the partial MB distributions defined in Sec.~\ref{ssec:approxmb}.
This is equivalent to keeping some of the amplitude bit-levels uniform and independent of the others.
The number of shaped and uniform amplitude bit-levels are denoted by $s$ and $u$, respectively, where $m-1=s+u$.

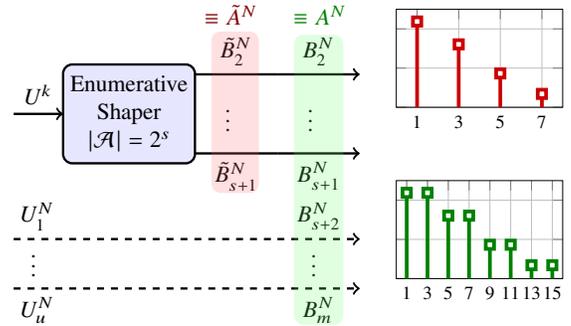
\begin{figure}[ht]
\hspace{0.8cm}
\resizebox{0.6\columnwidth}{!}{
\begin{tikzpicture}[line width=1pt]
\setlength{\hsep}{16ex} 
\draw [->] (0,0) -- (0.75,0) node[pos=0.5,above] {$U^k$};
\draw[rounded corners, fill=blue!10!white]  (0.75,-0.75) rectangle (2.75,0.75);
\node [align=center]  at (1.75,0) {Enumerative\\Shaper\\$\left|\calA\right|=2^s$};
    \draw [-] (2.75,0.6) -- (4,0.6) node[pos=0.5,above] {$\tilde{B}^N_{2}$};
    \draw [->] (4,0.6) -- (5.25,0.6) node[pos=0.5,above] {$B^N_{2}$};
    \node [align=center]  at (3.25,0) {$\vdots$};
    \node [align=center]  at (4.5,0) {$\vdots$};
    \draw [-] (2.75,-0.6) -- (4,-0.6) node[pos=0.5,below] {$\tilde{B}^N_{s+1}$};
    \draw [->] (4,-0.6) -- (5.25,-0.6) node[pos=0.5,below] {$B^N_{s+1}$};
    \draw [-,dashed] (0,-2.65) -- (4,-2.65) node[pos=0.09,below] {$U_u^{N}$};
    \draw [-,dashed] (0,-1.9) -- (4,-1.9) node[pos=0.09,above] {$U_1^{N}$};
    \node [align=center]  at (0.3,-2.2) {$\vdots$};
    \node [align=center]  at (4.55,-2.2) {$\vdots$};
    \draw [->,dashed] (4,-2.65) -- (5.25,-2.65) node[pos=0.5,below] {$B_m^{N}$};
    \draw [->,dashed] (4,-1.9) -- (5.25,-1.9) node[pos=0.5,above] {$B_{s+2}^{N}$};
    \draw[rounded corners, draw=none, fill opacity=0.2, fill=red!60!white]  (3.0,1.25) rectangle (3.72,-1.2);
    \node [align=center]  at (3.3,+1.5) {\textcolor{red!50!black}{$\equiv \tilde{A}^N$}};
    \draw[rounded corners, draw=none, fill opacity=0.2, fill=green!60!white]  (4.25,-3.2) rectangle (5,+1.2);
    \node [align=center]  at (4.65,+1.5) {\textcolor{green!50!black}{$\equiv A^N$}}; 
\node[align=center] at (5.25,-2.5) {
\begin{axis}[every axis/.append style={font=\footnotesize},
width=0.45\FigureWidth,
height=0.45\FigureHeight,
xmin=0,
xmax=16,
ymin=0,
ymax=0.25,
grid=both,
xtick={1,3,5,7,9,11,13,15},
yticklabels={,,}
]
\addplot+[ycomb] [color=green!50!black,mark=square*,mark size=1.8pt, thick,mark options={fill=white,solid},ultra thick]
  table[row sep=crcr]{%
1 0.218885146989361 \\
3 0.218885146989361 \\
5 0.160576092667549 \\
7 0.160576092667549 \\
9 0.0864191664361979 \\
11 0.0864191664361979 \\
13 0.0341195939068915 \\
15 0.0341195939068915 \\
};\end{axis}};
\node[align=center] at (5.25,0.1) {
\begin{axis}[every axis/.append style={font=\footnotesize},
width=0.45\FigureWidth,
height=0.45\FigureHeight,
xmin=0,
xmax=8,
ymin=0,
ymax=0.5,
grid=both,
xtick={1,3,5,7,9,11,13,15},
yticklabels={,,}
]
\addplot+[ycomb] [color=red!80!black,mark=square*,mark size=1.8pt, thick,mark options={fill=white,solid},ultra thick]
  table[row sep=crcr]{%
1 0.437770293978723\\
3 0.321152185335099\\
5 0.172838332872396\\
7 0.0682391878137831\\
};\end{axis}};
\end{tikzpicture}}
\caption{(Left) Partial-ESS block that realizes $s$-bit shaping for $s<m-1$. 
This block is to be used as the amplitude shaper in the PAS architecture, see Fig.~\ref{fig:PASblockdiag}. (Right) Exemplary output distributions of the enumerative shaper (red) and the overall partial ESS block (green) for $k/N=1.75$ with $s=2$ and $m=4$.}
\label{fig:PESSblock}
\end{figure}

We propose to use an enumerative shaper that works with the $2^{s+1}$-ASK amplitude alphabet, see Fig.~\ref{fig:PESSblock}.
This shaper maps $k$-bit message indices $U^k$ to shaped amplitude sequences $\tilde{A}^N$.
The distribution of $\tilde{A}^N$ is Gaussian-like over $\{1, 3,\cdots, 2^s-1\}$.
Accordingly, the corresponding binary amplitude label sequences $\tilde{B}_2^N\tilde{B}_3^N\cdots\tilde{B}_{s+1}^N$ are also shaped.
Thus, we now have $s$ shaped bit-levels which are highlighted by light red in Fig.~\ref{fig:PESSblock}.

Then $u$ additional $N$-bit data sequences $U_1^NU_2^N\cdots U_u^N$ are used as the uniform amplitude bit-levels for $2^m$-ASK and combined with the $s$ shaped levels outputted by the shaper, see the labels highlighted by light green in Fig.~\ref{fig:PESSblock}. 
The way uniform and shaped amplitude bit-levels are combined depends on the employed binary labeling strategies at the output of the shaper and in the symbol mapper, see Fig.~\ref{fig:PASblockdiag}.
In this work, we consider BRGCs.
Thus, we connect the extra uniform data sequences to the last $u$ amplitude bit-levels of $2^m$-ASK.
Shaped bit levels of $2^{s+1}$-ASK are connected to the bit-levels of $2^m$-ASK with the same index, see Fig.~\ref{fig:PESSblock}.
We give the following example to clarify this construction.

\subsubsection*{\bf Example (2-bit shaped 16-ASK)}
We consider a transmission scheme based on 16-ASK, i.e., $m=4$.
To have an $s=2$-bit shaped output distribution, an enumerative shaper employing $2^{s+1}=8$-ASK amplitude alphabet is used.
Outputs $\tilde{A}^N$ of this shaper are then amplitude labeled with $\tilde{B}_2^N\tilde{B}_3^N$ using the mapping
\begin{eqnarray}
1 \rightarrow 00,\hspace{0.5cm}3 \rightarrow 01,\hspace{0.5cm}5 \rightarrow 11,\hspace{0.5cm}7 \rightarrow 10.
\end{eqnarray}
We then use these bit levels $\tilde{B}_2^N\tilde{B}_3^N$ as the amplitude bit levels $B_2^NB_3^N$ of 16-ASK. 
Next, each label is concatenated with a uniform data bit and the result is outputted as the label of an amplitude $A$ from 16-ASK alphabet, see Table~\ref{tab:16askBRGC}.
Note that the shaped bits are $B_2B_3$ and the uniform bit is $B_4$ in this setting.
The types of the distributions of $\tilde{A}$ and $A$ at the outputs of the enumerative shaper and the overall P-ESS block are exemplified in Fig.~\ref{fig:PESSblock} by red and green plots, respectively.  

\subsubsection*{\bf Remark}
PDM can also be used to shape a subset of the amplitude bits~\cite{SteinerSB2018_PDM}.
The difference is that $P_X(x)$ is constrained to be a product distribution.
In~\cite{SteinerSB2018_PDM}, the bit-level distributions are optimized such that $\exp[X^2]$ is minimized.

\subsection{Implementation Aspects of P-ESS} 
The set of energies of the amplitudes from an $M/2$-ASK alphabet is $\calE = \{e_1, e_2, \cdots, e_{M/4} \}$ where $e_i = (2i-1)^2$.
Based on the first-level approximation proposed for $M$-ASK in Sec.~\ref{ssec:approxmb}, we define the set of average energies of the symbol pairs as $\calE_1 = \{e_{1,1}, e_{2,1},\cdots, e_{M/4,1}\}$ where $e_{j,1} = (1/2)\cdot\{(4j-3)^2+(4j-1)^2\}$ noting that $e_{j,1}=\mathbb{E}[|\calA_j|^2]$.
It is then by definition that $e_{l,1} = 4\cdot e_l +1$ for $l = 1,2,\cdots, M/4$.
This observation has two consequences:
\begin{itemize}
\item The bounded-energy ESS trellises constructed based on $\calE$ and $\calE_1$ have the same structure, i.e., the connections relating two consecutive columns, see Fig.~\ref{esstrellis}.
\item The MB distribution over \small{$\sqrt{\calE}$} for $i = 1, 2,\cdots,M/4$ and the MB distribution over $\sqrt{\calE_1}$ for $j = 1, 2,\cdots,M/4$ are the same given that they have the same entropy, where $\sqrt{\calE}$ indicates the set of square roots of elements in $\calE$.
\end{itemize}

\subsection{PAS for Lower Code Rates} 
In the PAS scheme, there is a lower bound on the FEC code rate that is $R_c \geq (m-1)/m$~\cite[Sec. IV]{BochererSS2015_ProbAmpShap}.
This is due to the fact that by prescribing the amplitudes at the output of the shaper, $(m-1)$~bit/1-D are already fixed prior to FEC coding, see Fig.~\ref{fig:PASblockdiag}.
Thus the encoder can at most add 1 bit redundancy per symbol making the smallest possible code rate $(m-1)/m$. 
However when P-ESS is used, only $s<m-1$ of the amplitude bit/1-D are fixed by the shaping process.
Accordingly, instead of using information bits for the remaining $u$ bit-levels as in Sec.~\ref{ssec:pess}, we can use the parity added by the encoder.
Thus, we can relax the code rate constraint to $R_c>s/m$.

\section{Results and Discussion}\label{sec:results}
\subsection{Rate Loss Results} 
\begin{figure}[ht]
\centering
\includegraphics[width=\columnwidth]{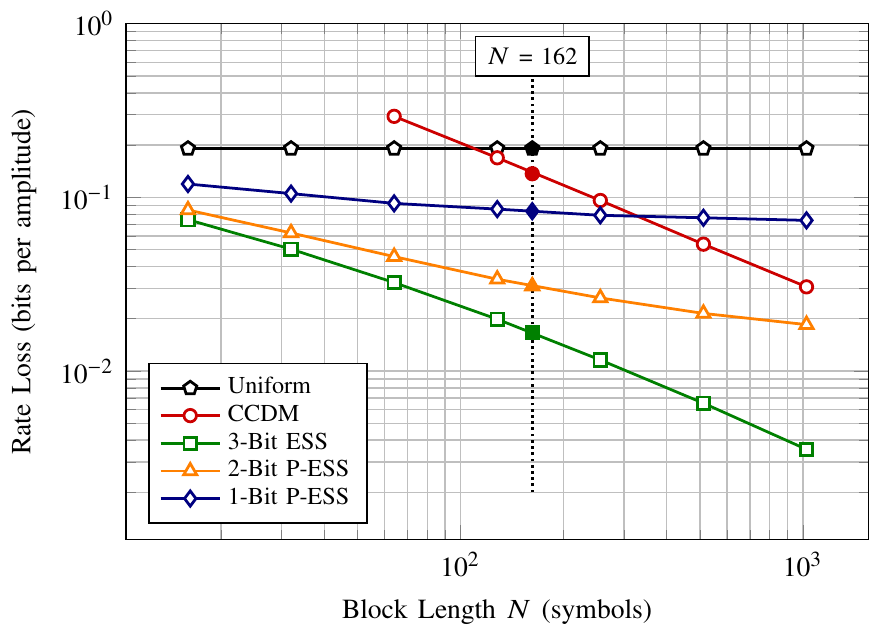}
\caption{$\rloss$ vs. block length of 16-ASK for various shaping schemes.
}
\label{rloss}
\end{figure}

Figure~\ref{rloss} shows $\rloss$ in \eqref{eq:rloss} vs. $N$ for 1- and 2-bit P-ESS, 3-bit ESS and CCDM~\cite{SchulteB2016_CCDM}.
For comparison, the same is plotted also for uniform signaling.
The target shaping rate is $R_s=2.6667$ with 16-ASK.
As some of the amplitude bits are kept uniform and independent of the others, $\rloss$ converges to a non-zero value for 1- and 2-bit P-ESS, i.e., 0.071 and 0.015 bit/amp., respectively, unlike the 3-bit shaped schemes.
However, for this example, CCDM requires roughly $N>300$ to surpass 1-bit P-ESS.
This shows that shaping some amplitude bits using ESS provides a better rate loss performance than CCDM in the short block length regime.
We note here that the rate losses of ESS and CCDM also depend on the constellation size.

\subsection{End-to-End Decoding Results}
Monte Carlo simulation is used to evaluate the frame error rate (FER) performance of the proposed P-ESS technique in the PAS framework for 16-ASK.
The binary labeling given in Table~\ref{tab:16askBRGC} is employed and combined with 648-bits long, systematic LDPC codes from IEEE 802.11~\cite{IEEE80211_2016}, leading to $N=162$.
All 2- and 1-bit P-ESS, and 3-bit ESS schemes are considered along with CCDM and uniform signaling.
The target rate is $R_t = 3$~bit/1-D.
Shaping techniques are coupled with the rate $R_c = 5/6$ code leading to $\gamma=1/3$ where the uniform transmission is with $R_c = 3/4$.
The rate of the amplitude shaping block is $k/N = R_t - \gamma = 2.667$.
Corresponding shaping parameters are tabulated in Table~\ref{tab:shaparam}\footnote{The notation $\#(a)$ is used to indicate the number of occurrences of $a$ in a constant composition $N$-sequence. Thus $\sum_{a\in\calA} \#(a) = N$~\cite{SchulteB2016_CCDM}.}.

\begin{table}[h]
\renewcommand{\arraystretch}{1.3}
\centering
\caption{Shaping Parameters for $m=4$, $\gamma=1/3$ and $R_t = 3$~bit/1-D}
\begin{tabular}{cccccc} 
\hline
 Method & $u$ &$\emax$ or $\#(a)$ & $k/N$ & $E$ & $\Gs$\\
\hline\hline
ESS   & 0 & 6514 & 2.667 & 39.69 & 1.29\\
\hline
P-ESS & 1 & 1626 & 1.667 & 40.73 & 1.18\\
P-ESS & 2 & 402 & 0.667 & 44.44 & 0.81 \\
\hline
CCDM & 0 & (34, 32, 28, 23, 18, 13, 9, 5) & 2.667 & 48.31 & 0.44\\
\hline
\end{tabular}
\label{tab:shaparam}
\end{table}

\begin{figure}[ht]
\centering
\includegraphics[angle=90,width=0.92\columnwidth]{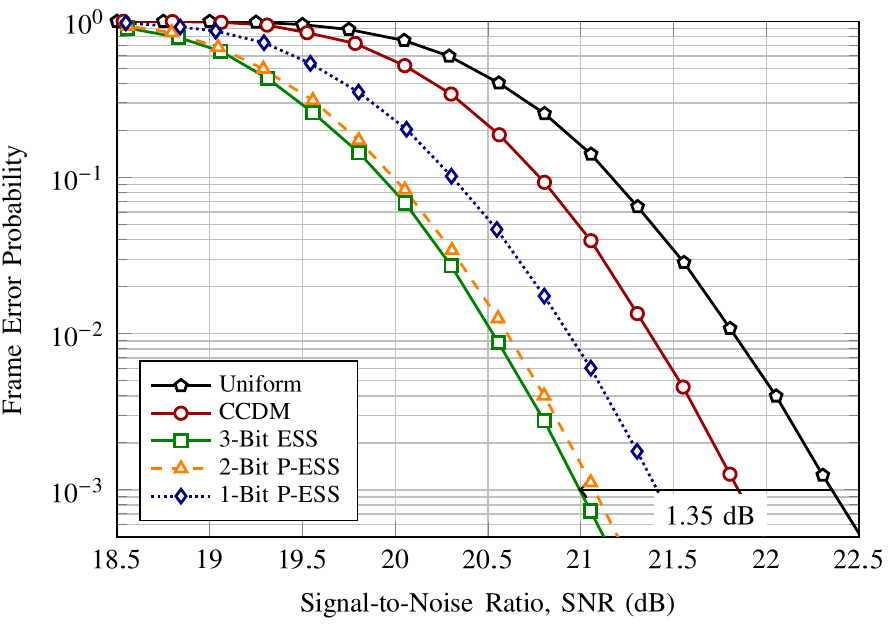}
\caption{FER vs. SNR for 16-ASK at a rate of 3 bit/1-D. IEEE 802.11 LDPC codes with 648 bits are used, i.e., $N=162$.
}
\label{fep162}
\end{figure}

The FER performance of the shaped and uniform schemes is shown in Fig.~\ref{fep162}.
We observe that at an FER of 10$^{-3}$, ESS performs 1.35 dB more energy-efficiently than uniform signaling.
Here the 2-bit and 1-bit P-ESS provide 1.27 and 0.95 dB improvement, respectively, while the gain is 0.45 dB for CCDM.
Firstly, all these values roughly match the corresponding shaping gains $\Gs$ given in Table~\ref{tab:shaparam} in dB.
Secondly, as claimed following the discussion in Sec.~\ref{ssec:approxmb}, the 2-bit P-ESS operates very close to the 3-bit ESS, i.e., in its 0.08 dB vicinity.
This provides operational evidence for our claim that not all amplitude bits have to be shaped to close most of the shaping gap.
We note that the relative performance of all techniques are as predicted by their rate losses in Fig.~\ref{rloss}.

Finally, required storage and computational complexity of the ESS-based schemes in Fig.~\ref{fep162} are tabulated in Table~\ref{tab:complexity}.
Here the bounded-precision ESS implementation is employed~\cite{GultekinWHS2018_ApproxEnumerative}.
We see that by shaping 2 amplitude bits instead of 3, the required storage and computational complexity of shaping can be decreased by factors of 6 and 3, respectively.
This reduction is accomplished in the expense of only 0.08 dB in performance, see Fig.~\ref{fep162}.
Although shaping just 1 amplitude bit provides limited gains, it brings a design flexibility enabling a trade-off between the shaping gain and shaping complexity.

\begin{table}[h]
\renewcommand{\arraystretch}{1.3}
\centering
\caption{}
\begin{tabular}{c||c|c} 
\hline
\begin{tabular}{@{}c@{}}{\bf Shaping} \\ {\bf Technique} \end{tabular} & \begin{tabular}{@{}c@{}}{\bf Storage (kilobytes)} \\ $L(N+1)(n_m+n_p)$~\cite{GultekinHKW2019_ESSforShortWlessComm}\end{tabular} & \begin{tabular}{@{}c@{}}{\bf Computation (bit oper./1-D)} \\ $(|\calA|-1)(n_m+n_p)$~\cite{GultekinHKW2019_ESSforShortWlessComm}\end{tabular}\\
\hline\hline
3-bit ESS   & 421.15 & 182 ($n_m=17$, $n_p=9$)  \\
2-bit P-ESS & 71.23  &  57 ($n_m=10$, $n_p=9$)  \\
1-bit P-ESS & 9.47   &  15 ($n_m=8$, $n_p=7$)   \\
\hline
\end{tabular}
\label{tab:complexity}
\end{table}

\section{Conclusion}
In this paper, we first defined an approximation for Maxwell-Boltzmann distributions and showed that the increase in gap-to-capacity that this causes is small.
This approximation suggests that some of the amplitude bit-levels may be kept uniform and independent of the others.
Then we proposed partial enumerative sphere shaping (P-ESS) to shape a subset of the amplitude bit-levels in the probabilistic amplitude shaping (PAS) framework.
This so-called P-ESS has lower rate loss than CCDM at a shaping rate of 2.67 bit/amp. using 16-ASK for block lengths smaller than 300.
Simulations over the AWGN channel demonstrate that shaping two amplitude bits of 16-ASK provides similar gains to shaping three bits, i.e., 1.3 dB over uniform signaling at rate 3~bit/1-D, with significantly smaller required storage and computational complexity.

\bibliographystyle{IEEEtran}
\bibliography{IEEEabrv,REFERENCES.bib}

\begin{thebibliography}{10}
\providecommand{\url}[1]{#1}
\csname url@samestyle\endcsname
\providecommand{\newblock}{\relax}
\providecommand{\bibinfo}[2]{#2}
\providecommand{\BIBentrySTDinterwordspacing}{\spaceskip=0pt\relax}
\providecommand{\BIBentryALTinterwordstretchfactor}{4}
\providecommand{\BIBentryALTinterwordspacing}{\spaceskip=\fontdimen2\font plus
\BIBentryALTinterwordstretchfactor\fontdimen3\font minus
  \fontdimen4\font\relax}
\providecommand{\BIBforeignlanguage}[2]{{%
\expandafter\ifx\csname l@#1\endcsname\relax
\typeout{** WARNING: IEEEtran.bst: No hyphenation pattern has been}%
\typeout{** loaded for the language `#1'. Using the pattern for}%
\typeout{** the default language instead.}%
\else
\language=\csname l@#1\endcsname
\fi
#2}}
\providecommand{\BIBdecl}{\relax}
\BIBdecl

\bibitem{BochererSS2015_ProbAmpShap}
G.~{B\"{o}cherer}, F.~{Steiner}, and P.~{Schulte}, ``Bandwidth efficient and
  rate-matched low-density parity-check coded modulation,'' \emph{IEEE Trans.
  Commun.}, vol.~63, no.~12, pp. 4651--4665, Dec 2015.

\bibitem{SchulteB2016_CCDM}
P.~Schulte and G.~B\"{o}cherer, ``Constant composition distribution matching,''
  \emph{IEEE Trans. Inf. Theory}, vol.~62, no.~1, pp. 430--434, Jan. 2016.

\bibitem{Fehenberger2019_MPDM}
T.~{Fehenberger}, D.~S. {Millar}, T.~{Koike-Akino}, K.~{Kojima}, and
  K.~{Parsons}, ``Multiset-partition distribution matching,'' \emph{IEEE Trans.
  on Commun.}, vol.~67, no.~3, pp. 1885--1893, Mar. 2019.

\bibitem{WillemsW1993_ESS}
F.~Willems and J.~Wuijts, ``A pragmatic approach to shaped coded modulation,''
  in \emph{Proc. Symp. on Commun. and Veh. Technol. in the Benelux}, Oct. 1993.

\bibitem{Schulte2019_SMDM}
P.~{Schulte} and F.~{Steiner}, ``Divergence-optimal fixed-to-fixed length
  distribution matching with shell mapping,'' \emph{IEEE Wireless Commun.
  Lett.}, vol.~8, no.~2, pp. 620--623, Apr. 2019.

\bibitem{Forney1984_EffModBLchan}
G.~Forney, R.~Gallager, G.~Lang, F.~Longstaff, and S.~Qureshi, ``Efficient
  modulation for band-limited channels,'' \emph{IEEE J. Sel. Areas Commun.},
  vol.~2, no.~5, pp. 632--647, Sep. 1984.

\bibitem{Gultekin2018_ConstShapforShortBlocks}
Y.~C. G\"{u}ltekin, W.~J. van Houtum, and F.~M.~J. Willems, ``On constellation
  shaping for short block lengths,'' in \emph{Proc. Symp. on Inf. Theory and
  Signal Process. in the Benelux (SITB)}, Enschede, The Netherlands, June 2018,
  pp. 86--96.

\bibitem{GultekinHKW2019_ESSforShortWlessComm}
Y.~C. G\"{u}ltekin, W.~J. van Houtum, A.~Koppelaar, and F.~M. Willems,
  ``{Enumerative Sphere Shaping for Wireless Communications with Short
  Packets},'' \emph{IEEE Trans. Wireless Commun.}, Oct. 2019.

\bibitem{SteinerSB2018_PDM}
F.~Steiner, P.~Schulte, and G.~B\"{o}cherer, ``Approaching waterfilling
  capacity of parallel channels by higher order modulation and probabilistic
  amplitude shaping,'' in \emph{Proc. Conf. on Inf. Syst. and Sci.}, Princeton,
  NJ, U.S.A., Mar. 2018.

\bibitem{CoverT1991_ElementsofInfoTheo}
T.~M. Cover and J.~A. Thomas, \emph{Elements of Information Theory}.\hskip 1em
  plus 0.5em minus 0.4em\relax New York, NY, USA: John Wiley \& Sons, 1991.

\bibitem{Kschischang1993_OptimalNonUnifSignaling}
F.~R. Kschischang and S.~Pasupathy, ``Optimal nonuniform signaling for gaussian
  channels,'' \emph{IEEE Trans. Inf. Theory}, vol.~39, no.~3, pp. 913--929, May
  1993.

\bibitem{LaroiaFT1994_OptimalShaping}
R.~{Laroia}, N.~{Farvardin}, and S.~A. {Tretter}, ``On optimal shaping of
  multidimensional constellations,'' \emph{IEEE Trans. Inf. Theory}, vol.~40,
  no.~4, pp. 1044--1056, July 1994.

\bibitem{Bocherer2014_ProbSigShapForBMD}
G.~B\"{o}cherer, ``Probabilistic signal shaping for bit-metric decoding,'' in
  \emph{Proc. IEEE Int. Symp. Inf. Theory}, Honolulu, HI, USA, June 2014, pp.
  431--435.

\bibitem{Wachsmann1999_MLcodes}
U.~Wachsmann, R.~F.~H. Fischer, and J.~B. Huber, ``Multilevel codes:
  theoretical concepts and practical design rules,'' \emph{IEEE Trans. Inf.
  Theory}, vol.~45, no.~5, pp. 1361--1391, July 1999.

\bibitem{GultekinWHS2018_ApproxEnumerative}
Y.~C. {G\"{u}ltekin}, F.~M.~J. {Willems}, W.~J. {van Houtum}, and
  S.~{\c{S}erbetli}, ``Approximate enumerative sphere shaping,'' in \emph{Proc.
  IEEE Int. Symp. Inf. Theory}, Vail, CO, U.S.A., June 2018, pp. 676--680.

\bibitem{IEEE80211_2016}
\emph{IEEE Standard for Inform. Technol.-Telecommun. and Inform. Exchange
  Between Syst. Local and Metropolitan Area Networks-Specific Requirements-Part
  11: Wireless LAN Medium Access Control (MAC) and Physical Layer (PHY)
  Specifications}, {IEEE} Standard 802.11-2016 (Revision of IEEE Standard
  802.11-2012), Dec. 2016.

\end{thebibliography}

\end{document}